# Temperature-dependent multi-k magnetic structure in multiferroic $Co_3TeO_6$

S. A. Ivanov[ab], R. Tellgren[c], C. Ritter[d], P. Nordblad[b], R. Mathieu[b*],
G. André[e], N.V. Golubko[a], E.D. Politova[a], M. Weil[f]

a- Department of Inorganic Materials, Karpov' Institute of Physical Chemistry, Vorontsovo pole, 10 105064, Moscow K-64, Russia
b- Department of Engineering Sciences, Uppsala University, Box 534, SE-751 21 Uppsala, Sweden
c- Department of Materials Chemistry, Uppsala University, Box 538, SE-751 21 Uppsala, Sweden
d- Institute Laue Langevin, Grenoble, France
e- Laboratoire Leon Brillouin, CEA, Saclay, France
f- Institute for Chemical Technologies and Analytics, Vienna University of Technology, A-1060 Vienna, Austria

**Abstract**
A complex magnetic order of the multiferroic compound $Co_3TeO_6$ has been revealed by neutron powder diffraction studies on ceramics and crushed single crystals. The compound adopts a monoclinic structure (s.g. *C2/c*) in the studied temperature range 2 K – 300 K but exhibits successive antiferromagnetic transitions at low temperature. Incommensurate antiferromagnetic order with the propagation vector $\mathbf{k}_1 = (0, 0.485, 0.055)$ sets in at 26 K. A transition to a second antiferromagnetic structure with $\mathbf{k}_2 = (0, 0, 0)$ takes place at 21.1 K. Moreover, a transition to a commensurate antiferromagnetic structure with $\mathbf{k}_3 = (0, 0.5, 0.25)$ occurs at 17.4 K. The magnetic structures have been determined by neutron powder diffraction using group theory analysis as a preliminary tool. Different coordinations of the $Co^{2+}$ ions involved in the low-symmetry *C2/c* structure of $Co_3TeO_6$ render the exchange-interaction network very complex by itself. The observed magnetic phase transformations are interpreted as an evidence of competing magnetic interactions. The temperature dependent changes in the magnetic structure, derived from refinements of high-resolution neutron data, are discussed and possible mechanisms connected with the spin reorientations are described.

## 1. Introduction

The origin and understanding of the coupling phenomena between different physical properties within a material is a central subject of solid state science. A great deal of theoretical and experimental attention in this field is currently focused on the coupling between magnetism and ferroelectricity, as can be encountered in the so-called multiferroics [1,2]. These compounds present opportunities for a wide range of potential applications [3,4] in addition to the fact that the fundamental physics of multiferroic materials is rich and fascinating [5-8]. The coexistence of ferromagnetism and ferroelectricity is difficult to achieve for many reasons [2, 9] and only very few multiferroic materials are known [5]. First, and most fundamentally, the cations which are responsible for the electric polarization in conventional ferroelectrics have filled *d*-electron shells. In contrast, ferromagnetism requires unpaired electrons, which in many materials are provided by *d* electrons of transition metal ions. Therefore the coexistence of the two phenomena, although not prohibited by any physical law or symmetry consideration, is discouraged by the local chemistry that favors one or the other but not both. In practice, alternative mechanisms for introducing both the polar ion displacements and the spin ordering are still needed. Theory gives us a good guide in what type of materials we can expect a large coupling effect. There are several important conclusions that we can draw regarding future directions of multiferroic research [2, 5, 9].

* Corresponding author.
*E-mail address:* roland.mathieu@angstrom.uu.se

However, experimental activity has been limited due to a lack of novel materials. Existing multiferroic compounds belong to different crystallographic classes, and although some general rules governing their behaviour are already established [10, 11], there is as yet no complete or general understanding of the origin of multiferroic behaviour.

With a few exceptions, the multiferroic materials that have been investigated are transition metal oxides, mainly with a perovskite-related structure. The unique range of responses, coupled with the flexibility of perovskites in accommodating a broad spectrum of atomic substitutions, provides a robust platform for probing correlations between structure, bulk chemistry and magnetoelectric properties [12-15].

The search for other structural families in which one can expect multiferroic behaviour including different types of magnetic and electric interactions which are responsible for the magnetic and ferroelectric properties and the magnetoelectric coupling should, however, not be neglected and the construction of new inorganic compounds through the rational combination of different cations has become an area of great interest. In this search for potential multiferroics, it is important to know how and why both magnetic and electric ordering emerges in a single phase. The crystal chemistry does not have answers to these questions, however, it is capable to determine under which conditions this might happen and find the compounds with these conditions fulfilled.

Despite intensive studies correlating the magnetic state of the material with its crystal chemical characteristics, the prime achievement in this field still is the discovery of the Goodenough–Kanamori–Anderson rules for the determination of the sign and magnitude of exchange in insulators, which was accomplished as far back as in the 1950–1960s [16–18]. According to these rules, a linear cation–anion–cation interaction (M–X–M) between half-filled orbitals is antiferromagnetic (AF), whereas cation–anion–cation interaction at less than 90° between half-filled orbitals is ferromagnetic (FM), provided the orbitals are connected with orthogonal anion orbitals. The superexchange including the $\sigma$-bonds is stronger than the superexchange for $\pi$-bonds. Among the cations with same electronic configuration, the superexchange of the cations with high valency is stronger. These rules are widely used; however the limits of their application are restricted since they only predict the ordering type between nearest neighbors.

Recently, complex metal oxides $A_3TeO_6$ (ATO) have begun to garner renewed interest in the research community due to their low temperature magnetic properties [19-22]. These materials are known to exhibit diverse properties as a result of compositional flexibility, however, although ATOs were studied in the 1970s and 80s our understanding of the manipulation of this structure type is still poor compared to perovskites [23]. In the structures of several ATO tellurates it is possible to find some relationship with the perovskite structure; however, the presence of new crystallographically non-equivalent sites for magnetic cations provides extra degrees of freedom for the manipulation of the structure offering an additional compositional flexibility.

**2. Motivation**

The $Co_3TeO_6$ compound [24] belongs to a novel series of the complex metal oxides $A_3TeO_6$ that show interesting physical properties [19]. These compounds crystallize in the cryolite-related crystal structure, in which the magnetic ions occupy several different crystallographic positions. Correlations between these sublattices play the key role in the complex magnetic behaviour of these materials. Magnetisation, specific heat and neutron diffraction data show that CTO is antiferromagnetic at low temperatures exibiting subsequent magnetic phase transitions in the ordered state [19]. Many experimental and theoretical questions about these interesting systems still remain open [19-22]. Neutron powder diffraction studies may provide answers to some of them by determining the magnetic ground states of CTO. We report here the results of the first neutron diffraction experiments on the compound. The neutron powder diffraction experiments have been made on powder samples of CTO at different temperatures and it was found that this compound exhibits long-range incommensurate magnetic order in certain temperature ranges and commensurate order in other.

In order to shed more light on the complex magnetic behaviour of this compound ceramic and single crystal samples of CTO were made and supplementary magnetic, specific heat and high-resolution neutron diffraction measurements were performed. In this paper we present the magnetic structure and properties of CTO, which reveal remarkable manifestations of competing magnetic interaction.

## 3. Experimental

### 3.1 Sample preparation
#### 3.1.1 Single crystals
Single crystals of $Co_3TeO_6$ were grown by chemical transport reactions [25] from CoO and $TeO_3$ as starting materials. CoO was prepared by heating $Co(NO_3)_2 \cdot 6H_2O$ in a platinum crucible at 1473 K for one hour and subsequent quenching to room temperature in an ice bath; $TeO_3$ was prepared by heating $H_6TeO_6$ at 673 K for 12 hours. A mixture of CoO and $TeO_3$ in the stoichiometric ratio 3:1 was thoroughly ground and, together with 10 mg $PtCl_2$, was placed in a silica ampoule which was evacuated, sealed, and heated in a temperature gradient 1098 →1028 K. At this temperature, $PtCl_2$ decomposes with release of $Cl_2$ which then serves as the actual transport agent. After 5 days the transport reaction was completed and dark blue to black crystals of $Co_3TeO_6$, mostly with a prismatic pinacoidal form and edge-lengths up to 5 mm, had formed in the colder part of the ampoule.

#### 3.1.2 Ceramic samples
A high quality ceramic sample of CTO was prepared by solid state reaction following the method described elsewhere [19]. High purity $Co(NO_3)_2 \cdot 6H_2O$ and telluric acid $H_2TeO_4 \cdot 2H_2O$ were used as starting materials. The raw materials were weighed in appropriate proportions for the $Co_3TeO_6$ formula. The homogenized stoichiometric mixtures were calcined at 770K for 7 hours, grounded into fine powders, pressed and annealed again several times with temperature interval 100 K up to 970 K with intermediate millings.

### 3.2. Chemical composition
The chemical composition and the homogeneity of the prepared crystals and ceramic samples were analyzed by energy-dispersive spectroscopy (EDS) using a JEOL 840A scanning electron microscope and INCA 4.07 (Oxford Instruments) software. The analyses performed on several particles showed that the concentration ratios of Co and Te are the stoichiometric ones within the instrumental resolution (0.05).
For the samples prepared for diffraction measurements the cation contents were determined by inductively coupled plasma atomic emission spectroscopy performed with an ARL Fisions 3410 spectrometer.

### 3.3. X-ray diffraction
Structural investigation of CTO single crystals was performed on a SMART Bruker diffractometer and the results are in a good agreement with earlier published data [24].
The phase identification and purity of the powder sample was checked from X-ray powder diffraction (XRPD) patterns obtained with a D-5000 diffractometer using Cu $K_\alpha$ radiation. The ceramic sample of CTO were crushed into powder in an agate mortar and suspended in ethanol. A Si substrate was covered with several drops of the resulting suspension, leaving randomly oriented crystallites after drying. The XRPD data for Rietveld analysis were collected at room temperature on a Bruker D8 Advance diffractometer (Vantec position-sensitive detector, Ge-monochromatized Cu $K_\alpha$ radiation, Bragg-Brentano geometry, DIFFRACT plus software) in the $2\theta$ range 10-152° with a step size of 0.02° (counting time 15 s per step). The slit system was selected to ensure that the X-ray beam was completely within the sample for all $2\theta$ angles.

### 3.4. Second harmonic generation (SHG) measurements.
The material was characterized by SHG measurements in reflection geometry, using a pulsed Nd:YAG laser ($\lambda$=1.064 μm). The SHG signal $I_{2\omega}$ was measured from the polycrystalline sample relative to an α-quartz standard at room temperature in the Q-switching mode with a repetition rate of 4 Hz.

### 3.5. Magnetic measurements
The magnetization experiments were performed in a Quantum Design MPMSXL 5 T SQUID magnetometer. The magnetization (M) of single-crystal samples was recorded as a function of temperature T in the interval 5-300 K in 20 Oe field using zero-field-cooled (ZFC) and field-cooled (FC) protocols.

### 3.6. Specific heat measurements
Specific heat measurements were performed using a relaxation method between 2 K and 100 K on a Physical Properties Measurement System (PPMS6000) from Quantum Design Inc.

### 3.7. Thermogravimetric studies
The thermogravimetric studies were performed up to 900 K under air using a Perkin-Elmer TGA-DSC System 7 thermobalance.

### 3.8. Neutron powder diffraction
Because the neutron scattering lengths of Co and Te are different, the chemical composition can be observed by neutron powder diffraction (NPD) with good precision. The neutron scattering length of oxygen is comparable to those of the cations and NPD provides accurate information on its position and stoichiometry.

Two sets of NPD data were collected on different samples and with different emphasis. The D1A data sets were collected on CTO ceramic samples at the Institute Laue-Langevin (Grenoble, France) on the high resolution powder diffractometer D1A (wavelength 1.91 Å) in the $2\theta$-range 10– 156.9° with a step size of 0.1°. The sample was inserted in a cylindrical vanadium container and a helium cryostat was used to collect data at several decisive temperatures in the temperature range 5-295 K and used for the main crystallographic refinements. The G 4-1 powder patterns were obtained from the G 4-1 diffractometer at LLB-CEA-Saclay on a powder sample which was prepared from CTO single crystals. This diffractometer is equipped with a pyrolitic graphite monochromator and a linear multidetector composed of 800 cells (BF3) separated by 0.1° covering an angular range of 80° ($2\theta$). The incident wavelength was $\lambda$ = 2.426Å. The NPD data were collected at low temperatures down to 1.6 K using a standard He cryostat and with much finer temperature steps to follow the magnetic phase transitions.

Nuclear and magnetic refinements were performed by the Rietveld method using the FULLPROF software [26]. The diffraction peaks were described by a pseudo-Voigt profile function, with a Lorentzian contribution to the Gaussian peak shape. A peak asymmetry correction was made for angles below 35° ($2\theta$). Background intensities were estimated by interpolating between up to 40 selected points (low temperature NPD experimental data) or described by a polynomial with six coefficients. An analysis of coordination polyhedra of cations was performed using the IVTON software [27]. The magnetic propagation vector was determined from the peak positions of the magnetic diffraction lines using the K-search software which is included in the FULLPROF refinement package [26]. Magnetic symmetry analysis was then done using the program BASIREPS [28]. The different allowed models for the magnetic structure were one by one tested against the measured data. Each structural model was refined to convergence, with the best result selected on the basis of agreement factors and stability of the refinement.

## 4. Results

According to the elemental analysis done on 20 different crystallites, the metal compositions of CTO ceramic and single crystals are $Co_{2.97(4)}Te_{1.03(3)}$ and $Co_{2.99(4)}Te_{1.01(3)}$, respectively if the sum of the cations is assumed to be 4. The oxygen content determined with an iodometric titration for the samples was determined as 5.98(3) and 5.99(3). All these values are very close to the expected ratios and permit to conclude that the sample compositions are the nominal ones. The microstructure of the obtained powders, observed by scanning electron microscopy, reveals a uniform grain distribution.

Second harmonic generation (SHG) measurements at room temperature for both powder and crystals gave a negative result, thus testifying that at this temperature the CTO compound probably possesses a centrosymmetric crystal structure. Notwithstanding, these samples could still be non-centrosymmetric, but at a level detectable only with sensitivities beyond $10^{-2}$ of quartz [29].

The low-magnetic field (20 Oe) ZFC and FC magnetization data (H applied along the *c*-direction) of CTO is shown together with the heat capacity data in Figure 1. As discussed in detail by Hudl et al. [19], the temperature dependence of the magnetization and heat capacity indicates an antiferromagnetic order below 26K, followed by a first-order-like transition near 18K (evidence by a sharp peak in the heat capacity and drop in magnetization). Hudl et al. also evidenced a magnetic field induced electrical polarization below 21K, suggesting a complex magnetic structure [19].

The first crystallographic characterization of the samples was performed by X-ray powder diffraction analysis at room temperature which showed that the prepared samples of CTO formed single phase powders. The room temperature XRD patterns could be indexed on the basis of a monoclinic unit cell with $a = 14.8113(4)$ Å, $b = 8.8394(3)$ Å, $c = 10.3589(4)$ Å, $\beta = 94.83(1)°$. These values are in a reasonable agreement with the lattice parameters obtained by Becker [24].

XRPD patterns could be successfully refined by the Rietveld method using s.g. *C2/c*. The Co-O and Te-O bond lengths calculated from the refined lattice parameters and atomic coordinates (see Tables 1, 2) are in close agreement with the original structural work. Furthermore, the corresponding bond valence sum calculations are consistent with the presence of $Co^{2+}$, $Te^{6+}$ and $O^{2-}$ ions. The final results from the Rietveld refinements with observed, calculated and difference plots for XRPD pattern of CTO at 295 K are shown in Figure 2.

### 4.1 Nuclear structure

The structural refinement of CTO was carried out using the high-resolution neutron powder diffraction pattern from D1A, recorded at room temperature. The data were fitted in the *C2/c* space group using cell parameters reported for a single-crystal specimen by Becker [24] as starting point. The experimental, calculated, and difference neutron powder-diffraction profiles are shown in Figure 3a. There is no significant mismatch between the observed intensity and the calculated profile. The structural parameters and the reliability factors of the refinement are summarized in Table 1 and the main interatomic distances for CTO are listed in Table 2.

The $Co_3TeO_6$ structure contains five different Co and two different Te cations. A projection on the *ac*-plane of the structure is shown in Figure 4a. Two crystallographically distinct Te cations Te(1) and Te(2) occupy octahedral sites such that these polyhedra are not directly connected to each other. The Co polyhedra have corner, edge and face-sharing connections. Co cations can be regarded as being arranged in pseudo-hexagonal layers where four of the five crystallographically distinct Co cations (Co(1)-Co(4)) also occupy octahedral sites. In principal a coordination of Co(3) can be described as only five-coordinated inside a square-pyramid because the Co(3)-O(2) bond length is elongated owing to a face-sharing between Co(3) and Co(4) octahedra. Similar but smaller elongation was found also for Co(2) cations but for these cations a formal octahedral coordination is still valid. Co(5) cations sit in tetrahedral sites with unusual edge-sharing connections between two adjacent Co(5) tetrahedra leading to a very short Co-Co bond of about 2.65 Å.

The structure can be described in terms of a close packing of O atoms along [001], in which Te(1) and Te(2) occupy 1/6 of the octahedral voids in such a way that only isolated $TeO_6$ octahedra are present. Within double sheets of O two third of the octahedral voids are occupied by Co(l), Co(4) and Te(l) (see Figure 4b). These sheets alternate with two comparable sheets containing Te(2), Co(3) and Co(5), which occupy 1/2 of the octahedral voids( see Figure 4c). 1/12 of the tetrahedral voids are occupied by Co(2). So, altogether 5/9 of the octahedral and 1/9 of the tetrahedral voids are occupied. Oxygen atoms are approximately close packed in a mixed hexagonal-cubic *hhchhc* six-layer sequence along a axis where the two hexagonal layers contain O(1), O(2), O(3), O(5), O(6) and O(7) atoms, while the cubic layer contains O(4), O(8) and O(9) atoms.

**Magnetic structure**
As mentioned before the neutron diffraction pattern of CTO at room temperature is purely nuclear and can be refined in space group *C2/c* indicating the paramagnetic state at this temperature. The neutron diffraction patterns recorded on lowering the temperature revealed the appearance of additional magnetic peaks (see Figures 5 and 6).
The collected neutron powder diffraction patterns confirmed the existence of two sequential magnetic orderings, as it was claimed on the basis of magnetic susceptibility studies. Below $T_N = 26$ K, weak magnetic peaks occur. They cannot be indexed as simple multiples of the crystal unit cell, revealing an incommensurate (ICM) wave vector $\mathbf{k_1} = (0, 0.485(2), 0.055(3))$. The values of the two components $\mathbf{k_y}$ and $\mathbf{k_z}$ of the wave vector have been refined using the FULLPROF program.
The refined wave vector components for $\mathbf{k_1}$ magnetic structure show a deviation from the commensurate value for two components, the largest deviation being found for the z component while the deviation from the commensurateness of the y component is only modest. Fixing $\mathbf{k_y}$ and $\mathbf{k_z}$ components to their commensurate values leads to a 4% increase of $R_{wp}$-factor. This suggests that $\mathbf{k_1}$ magnetic structure is incommensurate with the crystal lattice. Needless to say, we cannot be specific about the actual magnetic structure of CTO in this temperature range on the basis of the few purely magnetic peaks and the presence of 5 independent Co-positions. Such details may only be answered using single crystal data not yet available to us.
A second set of additional magnetic peaks is present below 26 K, coexisting in a short temperature interval with the ICM peaks. Further temperature lowering leads to weakening of the ICM peaks and their transformation into diffuse scattering. At 17.4 K only the second set of magnetic peaks remains. This set can be indexed with the wave vector $\mathbf{k_2}=(0, 0, 0)$. Some of the new magnetic peaks overlap with the nuclear reflections, while others appear at the positions extinct in the paramagnetic pattern, as expected from an antiferromagnetic ordering.
Representation analysis implemented in the FULLPROF program has been used to determine the magnetic structure with $\mathbf{k_2}$. The Fourier coefficients describing possible spin configurations can be written as linear combinations of irreducible representations (IR) of the wave vector group (little group). The magnetic representations for the 4e and 8f sites, occupied by Co(1) and Co(2)-Co(5), can be decomposed in IRs:

$$\Gamma(4e) = \Gamma_1 + \Gamma_2 + 2\Gamma_3 + 2\Gamma_4$$
$$\Gamma(8f) = 3\Gamma_1 + 3\Gamma_2 + 3\Gamma_3 + 3\Gamma_4$$

Possible combinations of the magnetic moments are presented in Table 3. Refinement of the models has been performed with FULLPROF software. The best agreement with experimental patterns was obtained for model $\Gamma_4$. This antiferromagnetic ordering is visualized in Figure 7. The Co moments reach the values presented in Table 4. These values are smaller than the spin only component $3\mu_B$ of the $Co^{2+}$ ion confirming complete quenching of the orbital moment for all cobalt ions in this compound.
A closer inspection of pattern collected below $T_N=17.4$ K showed that several new magnetic peaks appeared (Figure 5), which also can be indexed using simple multiples of the crystallographic unit cell, thus implying that the magnetic structure is commensurate with the wave vector $\mathbf{k_3} \approx (0, 0.5, 0.25)$. The propagation vector of the $\mathbf{k_3}$-structure was determined using indexing programs like "K-search" which is part of the FULLPROF suite of programs. A complete group-theoretical analysis was performed as well for the $k_3$-structure, however, it did not give apart from the relation between sites coupled through the *C*-centering any constraints for the coupling of the magnetic moments on the different sites. This gives 18 independent Co-sites with each having allowed components in all three orthogonal directions. The refinement had therefore be performed using a "trial and error" method where we introduced, however, the strong constraint of fixing the components of the magnetic moments of Co-sites occupying the same sublattice to have the same value. There are three possibilities to couple 4 spins antiferromagneticaly: ++--, +--+ and +-+- . For the 2-fold site only the +- coupling is possible. As there are 3 independent orthogonal directions one has for every 4-fold site 27 different possibilities of coupling the spins. Without any additional constraint this gives for the four 4-fold sites about 500000 different possibilities. Trying the different couplings it became, however, quickly clear that only the ++-- type coupling is present (see Figure 8) as both the +--+ and the +-+- couplings created magnetic peak intensities at positions which had no intensity. Together with the fact that the *x*-component refined to about zero for all five different Co-sublattices the proposed magnetic structure is relatively simple as it needs only ten parameters to be described. The obtained structural model of commensurate antiferromagnetic structure can be seen in Table 5 and Figure 9. For some refinements the components of magnetic moments were found to be very small with relatively large standard deviations, indicating that moments do not exhibit these components. Finally, in the case of CTO, complex magnetic phase transitions have been observed, following the scheme summarized in Figure 10. The main feature of the studied magnetic structures of CTO is several Co sublattices. Given that there are five Co magnetic sublattices in CTO and clear magnetic contributions at *both* commensurate and incommensurate positions in the low temperature diffraction patterns we can associate one Co site with the commensurate magnetic order and the other site with the incommensurate magnetic order. The obvious question is now whether we can be specific as to which site is which. It is clear that Co(1) sublattice does not have a significant magnetic moment in $k_2$ magnetic structure. On the other hand this Co(1) site is clearly very strongly contributing to $k_3$ magnetic structure.

**Discussion**
It was found that the nuclear structure of CTO retains *C2/c* symmetry down to 1.6 K. In the performed NPD experiments at low temperatures no evidences were found for any Bragg reflections violating the C-centering. The evolution of the lattice constants at low temperatures for CTO is presented in Figure 11. The lattice contractions around 18 K are evident. Below this temperature, the lattice parameters change very little and no additional anomalies that would occur in an accompanying structural transition, are observed.
In fact, the geometrical parameters (distances, angles) determine the values of magnetic interactions to a great extent. Results from polyhedral analysis of the different cations in CTO at different temperatures are presented in Table 6. The first observation we can make is that the Co cations have shifted significantly (~0.6 Å) away from the centroid of its coordination polyhedron. Different variations of the Co-O distances were found. The five different Co sublattices are not crystallographically and magnetically equivalent, meaning that also the exchange fields are different. The Te cations have also moved away from the octahedral centers but these shifts are significantly smaller than Co-cation shifts. As a consequence the variations in Te-O distances are small. It was found that the polyhedral volume of the Te sites is smaller than for the Co sites.
It is clear that the main magnetic exchange pathways present in CTO may be connected with the shorter Co-Co distances range between 2.9 and 3.1 Å (Co(5)-Co(5) 2.92 A, Co(2)-Co(5) 2.97 A, Co(3)-Co(3) 3.10 A, Co(4)-Co(4) 3.10 A) and consequently direct interactions are not negligible (see Figure 12). On the other hand, several superexchange interactions Co-O-Co with mean exchange angles above of 110.0° give rise to dominating antiferromagnetic coupling. Below 26 K, CTO enters an incommensurate antiferromagnetically ordered state. Below 17.4 K the magnetic structure of CTO alters and is described by two different spin-structures: a commensurate and collinear one with $k_2$ = (0, 0, 0) and *C2/c* symmetry having the spins antiferromagnetically coupled within a unit cell and a commensurately modulated structure, being described by $k_3$=(0, 0.5, 0.25). The magnetic spins lie in the *b–c* plane, no significant magnetic moment is observed along the *a*-axis and all magnetic spins are perfectly antiferromagnetically compensated. The magnetic structure $k_3$ may be described by a sinusoidal modulation model where an average of a magnetic moment component varies sinusoidally with distance in the direction of the propagation vector **k**. Co spins change sign when going from one cell to the other along *b* direction, while the repeat period along the *c* axis is 4*c*.

The derived magnetic moments of the five chemically similar $Co^{2+}$ cations are markedly different (see Tables 4 and 5) and all values are well below the value 3 $\mu_B$, expected for a spin only moment for the isolated ions. The reason for these deviations may be different degree of geometrical frustration of the spins in different Co sublattices.

The results of this research should be followed by a theoretical approach as it shows that the competition between exchange interactions along various coupling pathways in this particular crystallographic system results in a variety of different magnetic structures.

**Conclusions**

$Co_3TeO_6$ adopts monoclinic C2/c cryolite-type structure at room temperature, and retains this symmetry down to 1.6 K. key features of the structure are five different Co and two different Te cations. The several short Co–Co interatomic separations are possibly responsible for the complex magnetic ordering of CTO.

Based on the neutron diffraction results it was possible to find several different magnetic phases, which conform with magnetization and specific heat results.

NPD experiments have shed new light on the magnetic state and confirm the existence of long range magnetic ordering in CTO compound below 26 K. The magnetic order is propagated with the incommensurate $k_1$ = (0, 0.485, 0.055) and commensurate $k_2$ = (0, 0, 0) and $k_3$ = (0, 0.5, 0.25) wave vectors. The set of several commensurate and incommensurate states and the magnetic phase transformations in CTO can be connected with a high degree of competing ferro- and antiferromagnetic interaction in the system.

**Acknowledgements**

Financial support of this research from the Swedish Research Council (VR), Göran Gustafsson Foundation and the Russian Foundation for Basic Research is gratefully acknowledged. We also gratefully acknowledge the support from N. Sadovskaya and S. Stefanovich during an EDS cation analysis and second harmonic generation testing, respectively.

**Table 1** Summary of the results of the structural refinements of the $Co_3TeO_6$ samples using XRPD and NPD data.

| Experiment | XRPD | NPD(ILL) | NPD(ILL) | NPD(LLB) |
|---|---|---|---|---|
| T,K | 295 | 295 | 5 | 1.6 |
| a[Å] | 14.8113(2) | 14.8107(4) | 14.7673(4) | 14.7832(6) |
| b[Å] | 8.8394(3) | 8.8334(5) | 8.8220(5) | 8.8396(4) |
| c[Å] | 10.3589(3) | 10.3577(4) | 10.3237(5) | 10.3428(6) |
| β,º | 94.834(6) | 94.849(7) | 94.915(7) | 95.032(9) |
| s.g. | *C2/c* | *C2/c* | *C2/c* | *C2/c* |
| **Co(1)** | | | | |
| $x$ | 0.5 | 0.5 | 0.5 | 0 |
| $y$ | -0.1848(6) | -0.1829(7) | -0.1955(7) | -0.1938(9) |
| $z$ | 0.25 | 0.25 | 0.25 | 0.25 |
| $B[Å]^2$ | 0.67(3) | 0.78(4) | 0.55(4) | 0.49(5) |
| **Co(2)** | | | | |
| $x$ | 0.8561(5) | 0.8611(6) | 0.8547(6) | 0.8522(9) |
| $y$ | -0.3556(4) | -0.3575(7) | -0.3562(7) | -0.3602(7) |
| $z$ | 0.2321(5) | 0.2301(8) | 0.2266(8) | 0.2218(8) |
| $B[Å]^2$ | 0.77(1) | 0.67(4) | 0.61(4) | 0.39(8) |
| **Co(3)** | | | | |
| $x$ | 0.5233(3) | 0.5187(6) | 0.5149(6) | 0.5139(9) |
| $y$ | -0.6526(4) | -0.6548(7) | -0.6693(7) | -0.7017(9) |
| $z$ | 0.0419(3) | 0.0401(8) | 0.0367(8) | 0.0700(8) |
| $B[Å]^2$ | 0.54(1) | 0.58(4) | 0.48(4) | 0.54(6) |
| **Co(4)** | | | | |
| $x$ | 0.6651(3) | 0.6704(6) | 0.6609(6) | 0.6712(8) |
| $y$ | -0.2931(4) | -0.2975(7) | -0.2956(7) | -0.2898(8) |
| $z$ | 0.0573(5) | 0.0579(8) | 0.0589(8) | 0.0513(9) |
| $B[Å]^2$ | 0.48(1) | 0.53(4) | 0.39(4) | 0.61(6) |
| Co(5) | | | | |
| $x$ | 0.8005(3) | 0.7987(6) | 0.8005(6) | 0.8091(7) |
| $y$ | -0.3621(4) | -0.3606(7) | -0.3658(7) | -0.3613(9) |
| $z$ | 0.5699(4) | 0.5672(8) | 0.5771(8) | 0.5665(9) |
| $B[Å]^2$ | 0.51(3) | 0.58(4) | 0.35(4) | 0.44(5) |
| Te(1) | | | | |
| $x$ | 0 | 0 | 0 | 0 |
| $y$ | 0.5 | 0.5 | 0.5 | 0.5 |
| $z$ | 0.5 | 0.5 | 0.5 | 0.5 |
| $B[Å]^2$ | 0.38(3) | 0.41(4) | 0.32(4) | 0.42(6) |
| Te(2) | | | | |
| $x$ | 0.6621(3) | 0.6634(6) | 0.6628(6) | 0.6637(8) |
| $y$ | -0.5011(2) | -0.4989(7) | -0.5014(7) | -0.4911(9) |
| $z$ | 0.2998(3) | 0.2982(8) | 0.3017(8) | 0.3006(9) |
| $B[Å]^2$ | 0.43(4) | 0.33(4) | 0.39(4) | 0.37(5) |
| O(1) | | | | |
| $x$ | 0.9348(4) | 0.9325(5) | 0.9285(5) | 0.9328(5) |
| $y$ | -0.3488(6) | -0.3415(4) | -0.3316(4) | -0.3393(7) |
| $z$ | 0.5703(5) | 0.5747(6) | 0.5629(6) | 0.5582(5) |
| $B[Å]^2$ | 0.97(4) | 0.89(5) | 1.06(5) | 0.79(5) |
| O(2) | | | | |
| $x$ | 0.6014(4) | 0.5981(5) | 0.5944(5) | 0.5937(5) |
| $y$ | -0.3418(6) | -0.3387(4) | -0.3415(4) | -0.3489(7) |
| $z$ | 0.1995(5) | 0.2021(6) | 0.2063(6) | 0.2021(5) |
| $B[Å]^2$ | 1.19(2) | 0.97(5) | 0.94(5) | 0.68(4) |
| O(3) | | | | |
| $x$ | 0.5986(4) | 0.5998(5) | 0.6001(5) | 0.5953(5) |
| $y$ | -0.6612(6) | -0.6631(4) | -0.6471(4) | -0.6318(7) |
| $z$ | 0.1915(5) | 0.1973(6) | 0.1900(6) | 0.2025(5) |
| $B[Å]^2$ | 0.97(2) | 0.88(5) | 0.86(5) | 0.73(4) |
| O(4) | | | | |
| $x$ | 0.7406(4) | 0.7478(5) | 0.7527(5) | 0.7555(5) |
| $y$ | -0.5237(6) | -0.5221(4) | -0.5228(4) | -0.5186(7) |
| $z$ | 0.6618(5) | 0.6693(6) | 0.6696(6) | 0.6598(5) |
| $B[Å]^2$ | 1.19(4) | 1.03(5) | 1.11(5) | 0.91(2) |
| O(5) | | | | |
| $x$ | 0.9314(4) | 0.9342(5) | 0.9289(5) | 0.9317(5) |
| $y$ | -0.5089(6) | -0.5111(4) | -0.5166(4) | -0.5076(7) |
| $z$ | 0.3347(5) | 0.3357(6) | 0.3366(6) | 0.3402(8) |
| $B[Å]^2$ | 0.98(5) | 1.01(5) | 0.84(5) | 0.81(5) |
| O(6) | | | | |
| $x$ | 0.5842(4) | 0.5894(5) | 0.5812(5) | 0.5879(5) |
| $y$ | -0.5137(6) | -0.5179(4) | -0.5205(4) | -0.5174(7) |
| $z$ | 0.4412(5) | 0.4389(6) | 0.4350(6) | 0.4306(5) |
| $B[Å]^2$ | 1.09(5) | 0.97(5) | 0.96(5) | 0.89(5) |
| O(7) | | | | |
| *x* | 0.9285(4) | 0.9293(5) | 0.9272(5) | 0.9285(5) |
| *y* | -0.6524(6) | -0.6576(4) | -0.6533(4) | -0.6416(7) |
| *z* | 0.5651(5) | 0.5608(6) | 0.5634(6) | 0.5598(5) |
| *B[Å]$^2$* | 0.92(4) | 0.87(5) | 0.81(5) | 0.77(5) |
| **O(8)** | | | | |
| *x* | 0.7407(4) | 0.7346(5) | 0.7382(5) | 0.7413(5) |
| *y* | -0.3403(6) | -0.3376(4) | -0.3392(4) | -0.3279(7) |
| *z* | 0.3914(5) | 0.3875(6) | 0.3946(6) | 0.4004(6) |
| *B[Å]$^2$* | 1.26(5) | 1.12(5) | 1.07(5) | 0.93(5) |
| **O(9)** | | | | |
| *x* | 0.7311(4) | 0.7289(5) | 0.7237(5) | 0.7189(5) |
| *y* | -0.6588(6) | -0.6515(4) | -0.6603(4) | -0.6518(7) |
| *z* | 0.3821(5) | 0.3902(6) | 0.3858(6) | 0.3898(7) |
| *B[Å]$^2$* | 1.04(4) | 0.92(5) | 0.91(5) | 0.81(5) |
| $R_p$,% | 1.34 | 2.41 | 4.81 | 3.10 |
| $R_{wp}$,% | 1.87 | 3.72 | 7.06 | 4.38 |
| $R_B$(%) | 6.18 | 4.19 | 5.24 | 2.96 |
| $R_{mag1}$(%) | - | - | 5.87 | 6.18 |
| $R_{mag2}$(%) | - | - | 18.4 | 14.5 |
| $\chi^2$ | 1.96 | 2.17 | 2.23 | 2.09 |

**Table 2.** Selected bond lengths [Å] of CTO samples.

| T,K | | 5 | 295 |
|---|---|---|---|
| **Te(1)** | O7 | 1.879(5) (x2) | 1.878(6) (x2) |
| | O5 | 1.914(4) (x2) | 1.934(6) (x2) |
| | O1 | 1.965(4) (x2) | 1.944(6) (x2) |
| **Te(2)** | O9 | 1.842(6) | 1.886(6) |
| | O3 | 1.912(4) | 1.890(6) |
| | O6 | 1.913(4) | 1.905(6) |
| | O2 | 1.952(6) | 1.927(7) |
| | O4 | 1.995(4) | 1.977(5) |
| | O8 | 2.006(5) | 1.984(6) |
| **Co(1)** | O2 | 1.979(6) (x2) | 2.002(6) (x2) |
| | O1 | 2.134(5) (x2) | 2.120(5) (x2) |
| | O5 | 2.140(4) (x2) | 2.170(4) (x2) |
| **Co(2)** | O4 | 1.898(6) | 1.947(6) |
| | O5 | 2.063(4) | 2.027(6) |
| | O7 | 2.074(4) | 2.043(6) |
| | O3 | 2.119(6) | 2.212(7) |
| | O9 | 2.333(4) | 2.408(5) |
| | O2 | 2.550(6) | 2.536(7) |
| **Co(3)** | O3 | 1.946(6) | 1.992(6) |
| | O6 | 1.973(4) | 2.005(6) |
| | O1 | 1.988(4) | 2.011(6) |
| | O7 | 2.065(6) | 2.019(7) |
| | O6 | 2.245(4) | 2.276(5) |
| | O2 | 2.865(6) | 2.910(7) |
| **Co(4)** | O2 | 1.924(6) | 1.884(6) |
| | O9 | 2.118(4) | 2.063(6) |
| | O9 | 2.119(4) | 2.114(6) |
| | O7 | 2.140(6) | 2.178(7) |
| | O2 | 2.323(4) | 2.273(5) |
| | O2 | 2.333(6) | 2.402(7) |
| **Co(5)** | O4 | 1.856(4) | 1.901(6) |
| | O2 | 1.927(6) | 1.944(7) |
| | O1 | 1.932(4) | 1.956(5) |
| | O2 | 2.038(5) | 2.017(7) |

**Table 3** Irreducible representations of the wave vector group for **k**=(0,0,0) in the space group *C2/c*. The notation Co(ij) was used with the index i=1-5 labeling the site of cations and index j=1-4 labeling the cations within the site. The coefficients u,v,w are fixed by the symmetry for one set but are independent for different sites.

| Cation | Site | $\Gamma_1$ | $\Gamma_2$ | $\Gamma_3$ | $\Gamma_4$ |
|---|---|---|---|---|---|
| Co11 | 4e | 0 u 0 | 0 u 0 | u 0 v | u 0 v |
| Co21 | | 0 u 0 | 0 -u 0 | u 0 v | -u 0 -v |
| | | | | | |
| Coi1 | 8f | u v w | u v w | u v w | u v w |
| Coi2 | i=2-5 | -u v -w | -u v -w | u -v w | u -v w |
| Coi3 | | u v w | -u -v -w | u v w | -u -v -w |
| Coi4 | | -u v -w | u -v w | u -v w | -u v -w |

**Table 4** Magnetic moments for magnetic structuire with **k**=(0,0,0) in the space group *C2/c*.

| Cation | Mx | My | Mz | $\mu_B$ |
|---|---|---|---|---|
| Co11 | -0.7(4) | 0 | 0.7(5) | 1.0(7) |
| Co12 | 0.7(4) | 0 | -0.7(5) | 1.0(7) |
| Co21 | 1.8(1) | 0.7(1) | 2.1(1) | 2.7(1) |
| Co22 | -1.8(1) | -0.7(1) | -2.1(1) | 2.7(1) |
| Co31 | -2.2(1) | 0.6(2) | -1.9(1) | 2.9(1) |
| Co32 | 2.2(1) | -0.6(2) | 1.9(1) | 2.9(1) |
| Co41 | 0.8(4) | 0.2(3) | 0.8(5) | 1.1(7) |
| Co42 | -0.8(4) | -0.2(3) | -0.8(5) | 1.1(7) |
| Co51 | -2.1(1) | 0.9 | -1.9(1) | 2.8(1) |
| Co52 | 2.1(1) | -0.9 | 1.9(1) | 2.8(1) |

**Table 5** Magnetic moments for magnetic structure with **k**=(0,0.5,0.25) in the space group *C2/c*.

| Cation | Mx | My | Mz | $\mu_B$ |
|---|---|---|---|---|
| Co(1) | 0 | 2.9(1) | 2.1(1) | 3.6(1) |
| Co(2) | 0 | 1.9(1) | -1.1(1) | 2.2(1) |
| Co(3) | 0 | 0.5(4) | 0.2(3) | 0.6(4) |
| Co(4) | 0 | -0.3(3) | -0.9(4) | 1.0(4) |

| | | | | | |
|---|---|---|---|---|---|
| Co(5) | 0 | -1.2(2) | 1.8(1) | 2.2(1) | |

**Table 6** Polyhedral analysis of $Co_3TeO_6$ at different temperatures (cn - coordination number, x – shift from centroid, ξ- average bond distance with a standard deviation,
V- polyhedral volume, ω- polyhedral volume distortion.

**T=295K s.g.C2/c**

| Cation | cn | x(Å) | ξ (Å) | V(Å³) | ω | Valence |
|---|---|---|---|---|---|---|
| **Co(1)** | 6 | 0.081 | 2.097+/-0.077 | 11.9(1) | 0.021 | 2.04 |
| **Co(2)** | 6 | 0.152 | 2.180+/-0.236 | 12.8(1) | 0.043 | 1.86 |
| **Co(3)** | 6 | 0.059 | 2.202+/-0.363 | 13.2(1) | 0.024 | 1.95 |
| **Co(4)** | 6 | 0.136 | 2.153+/-0.179 | 12.3(1) | 0.047 | 1.91 |
| **Co(5)** | 4 | 0.249 | 1.954+/-0.048 | 3.6(1) | 0.063 | 1.98 |
| **Te(1)** | 6 | 0 | 1.918+/-0.032 | 9.3(1) | 0.009 | 5.99 |
| **Te(2)** | 6 | 0.044 | 1.928+/-0.043 | 9.5(1) | 0.008 | 5.85 |

**T=5K s.g. C2/c**

| Cation | cn | x(Å) | ξ (Å) | V(Å³) | ω | Valence |
|---|---|---|---|---|---|---|
| **Co(1)** | 6 | 0.061 | 2.084+/-0.082 | 11.7(1) | 0.020 | 2.12 |
| **Co(2)** | 6 | 0.153 | 2.173+/-0.232 | 12.8(1) | 0.040 | 1.89 |
| **Co(3)** | 6 | 0.066 | 2.180+/-0.352 | 13.0(1) | 0.023 | 2.05 |
| **Co(4)** | 6 | 0.155 | 2.160+/-0.153 | 12.5(1) | 0.048 | 1.88 |
| **Co(5)** | 4 | 0.242 | 1.938+/-0.075 | 3.5(1) | 0.059 | 1.98 |
| **Te(1)** | 6 | 0 | 1.920+/-0.032 | 9.4(1) | 0.008 | 5.98 |
| **Te(2)** | 6 | 0.048 | 1.937+/-0.061 | 9.6(1) | 0.009 | 5.76 |

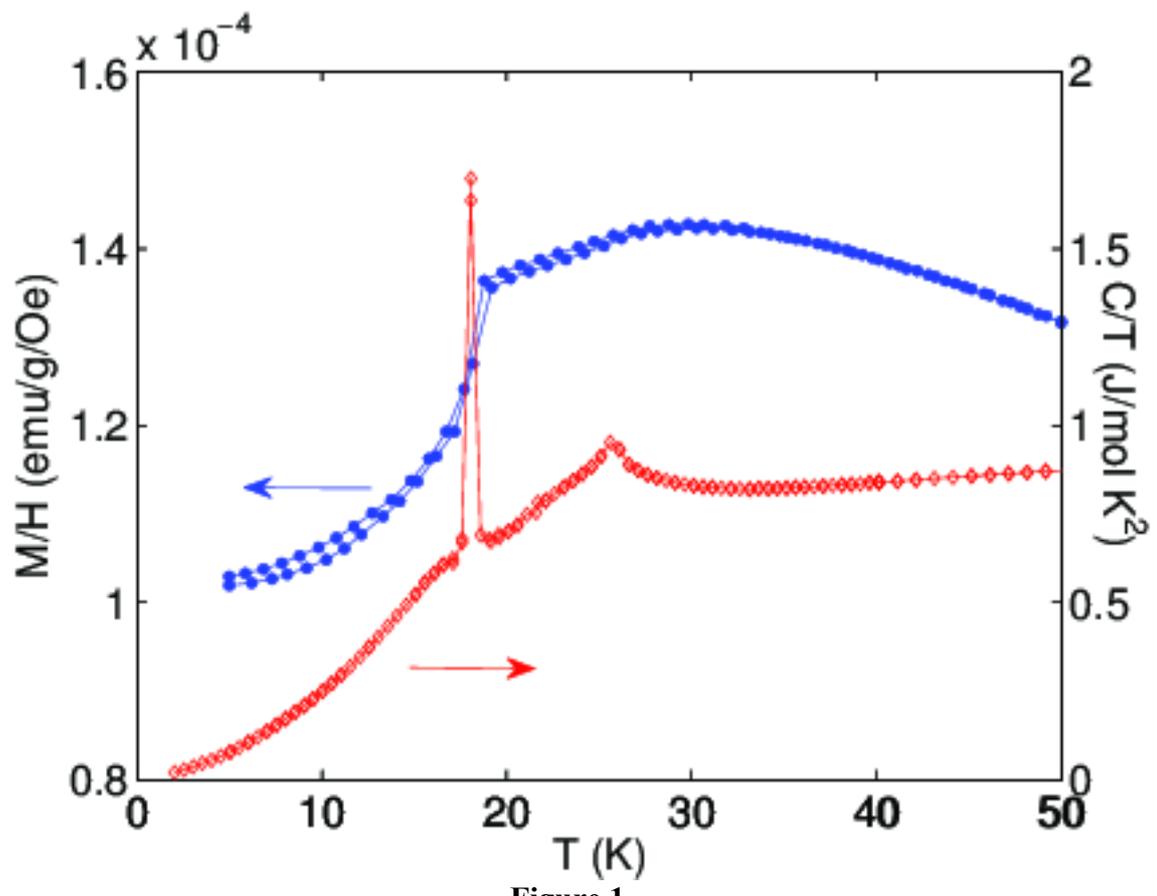

**Figure 1**

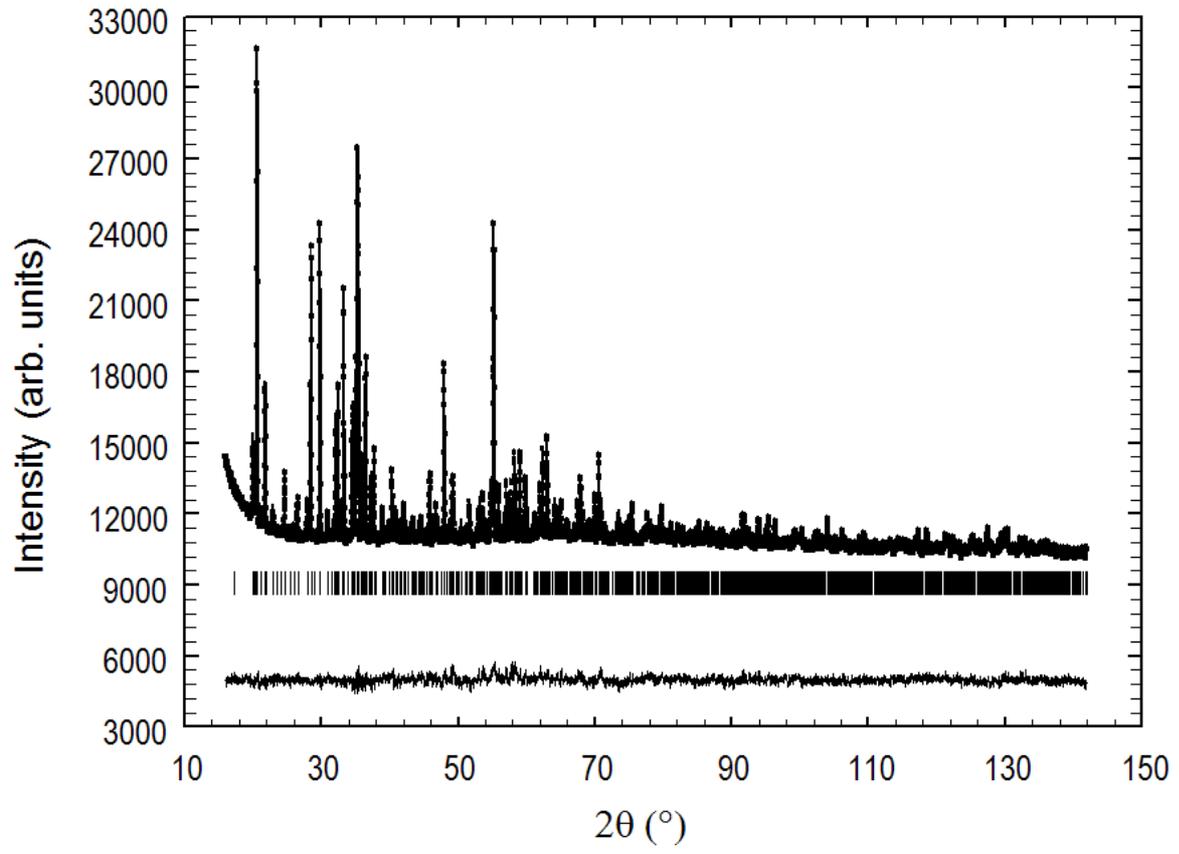

**Figure 2**

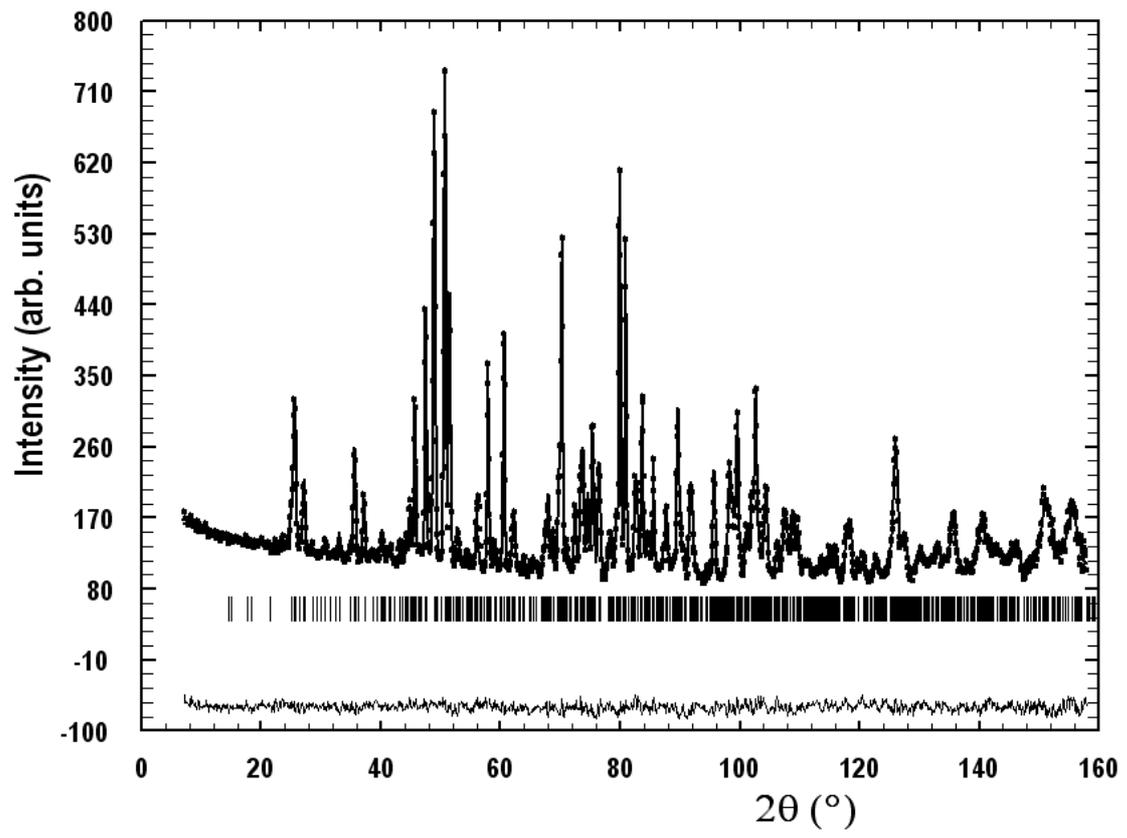

**Figure 3a**

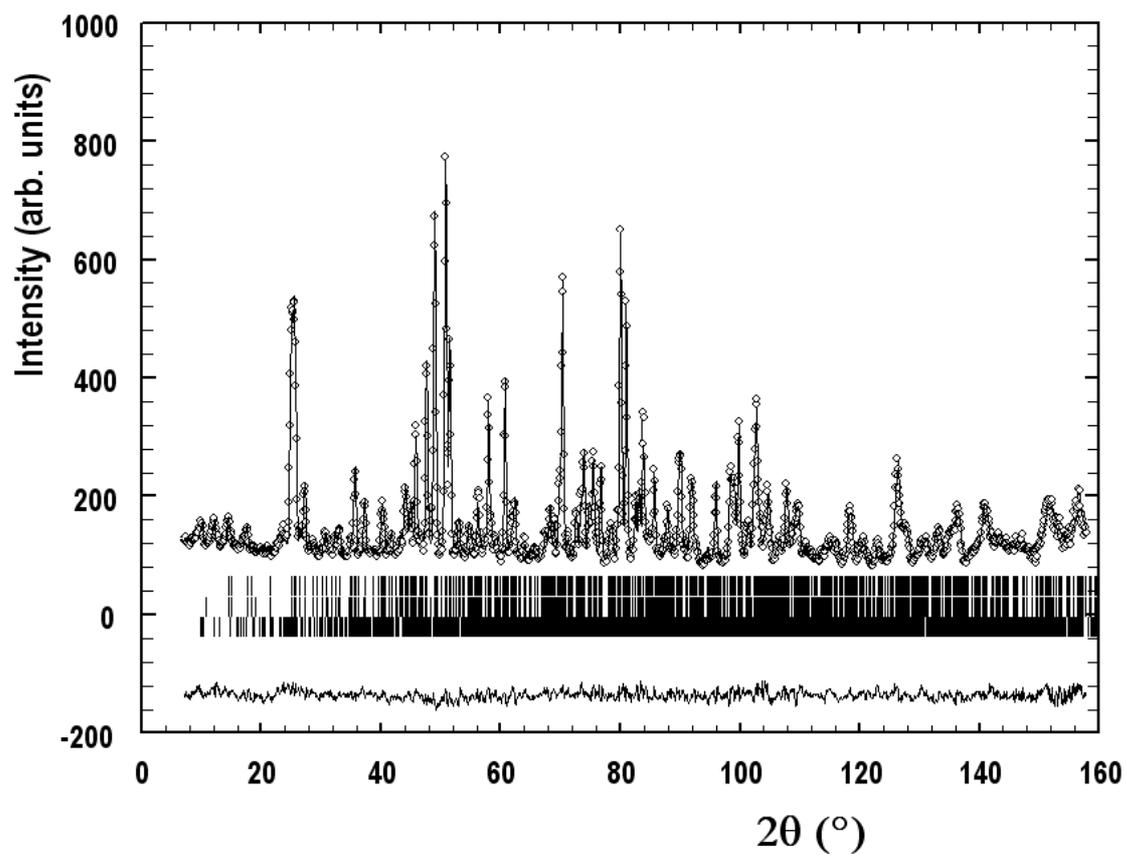

**Figure 3b**

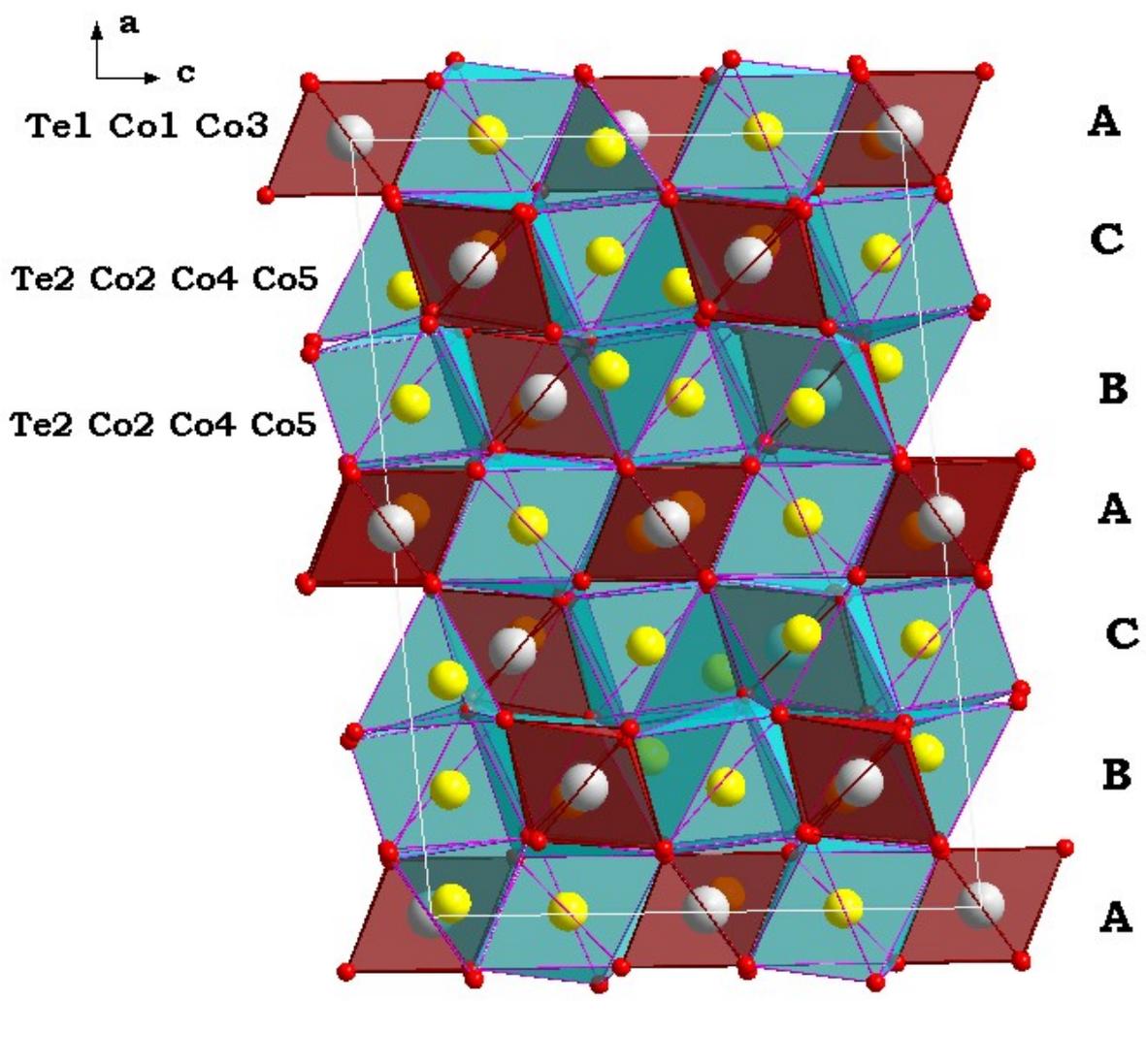

**Figure 4a**

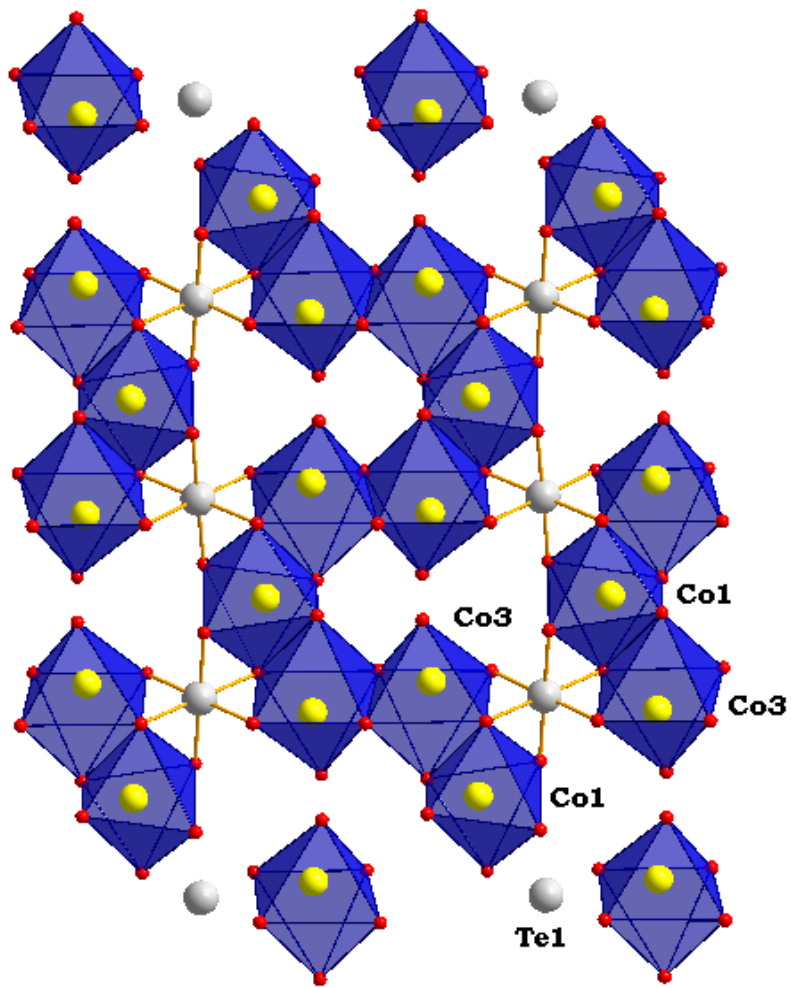

**Figure 4b**

**Figure 4c**

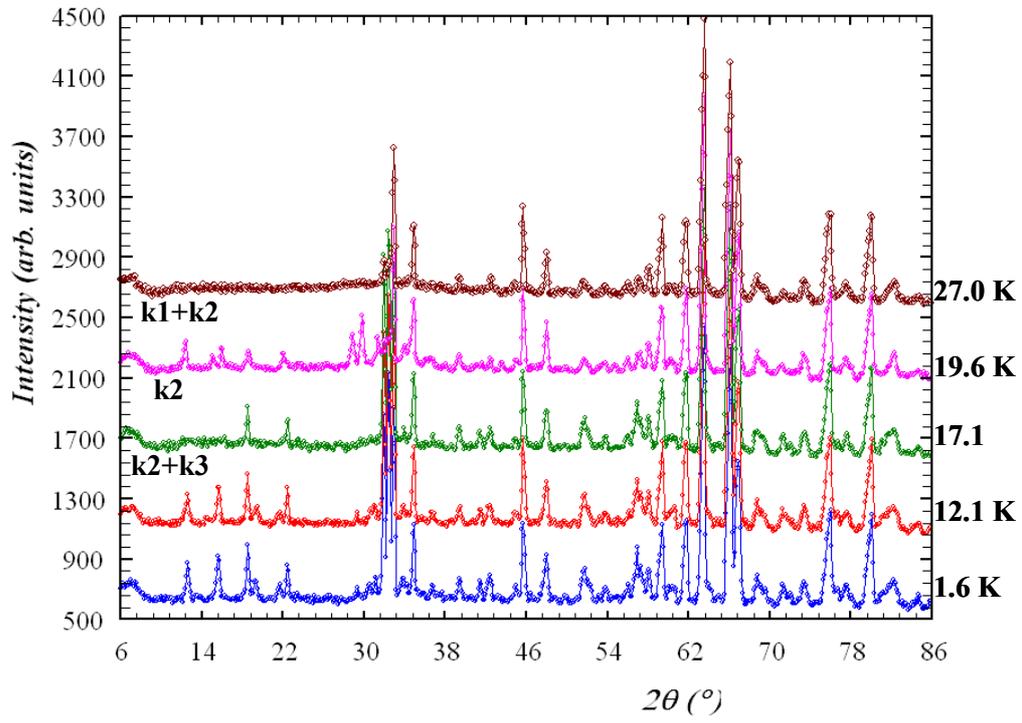

**Figure 5a**

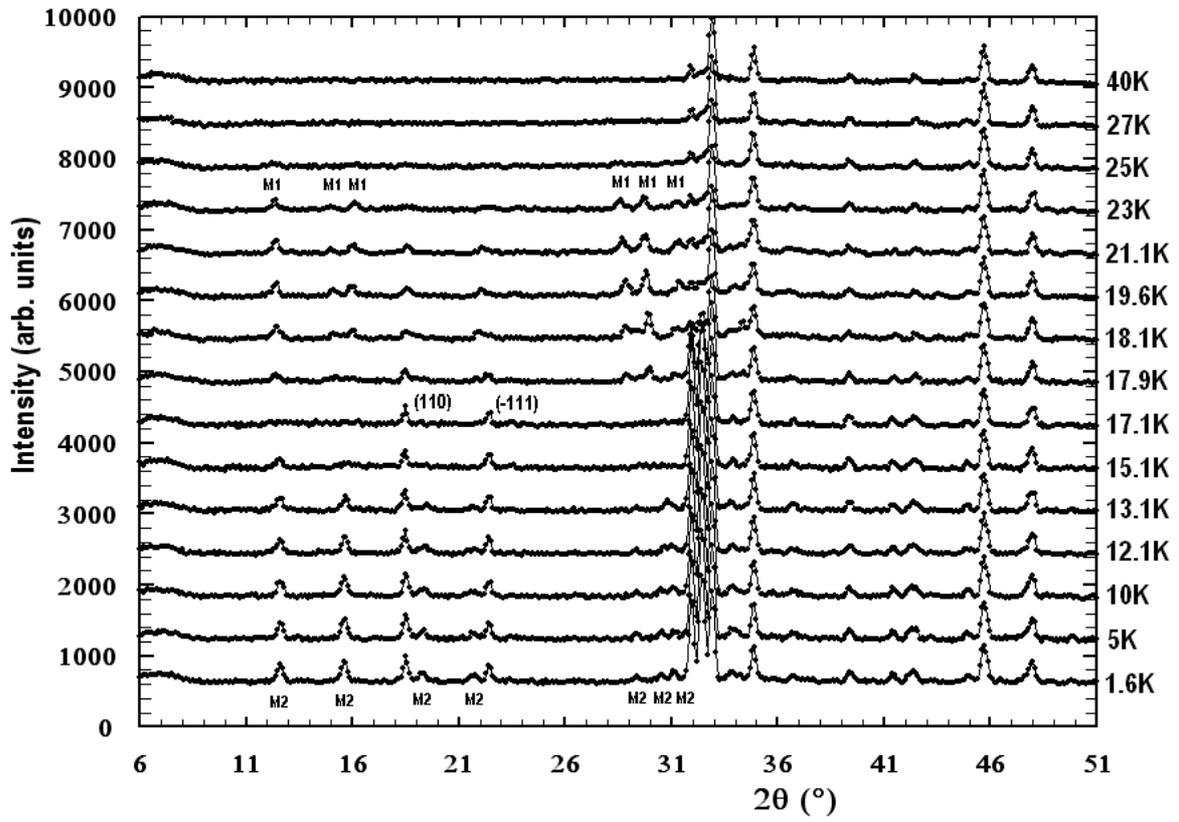

**Figure 5b**

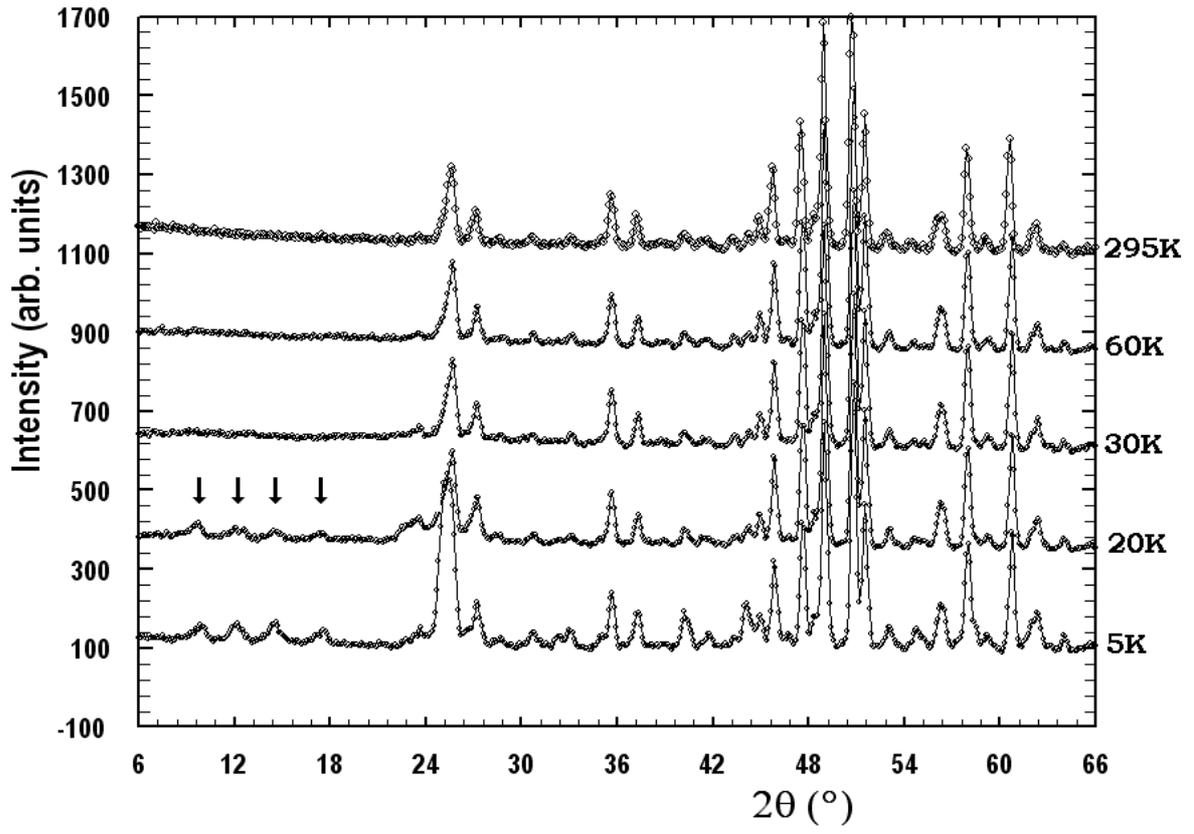

**Figure 6**

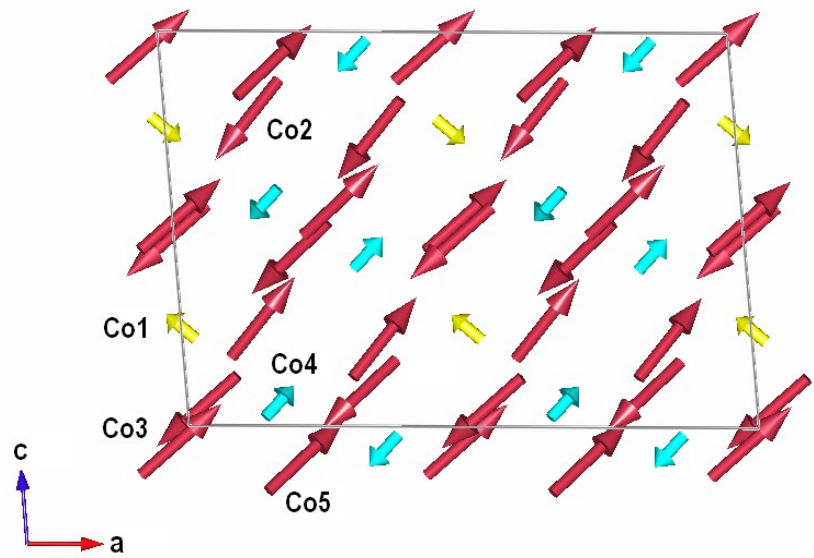

**Figure 7**

Co3TeO6 LLB 1.6 K

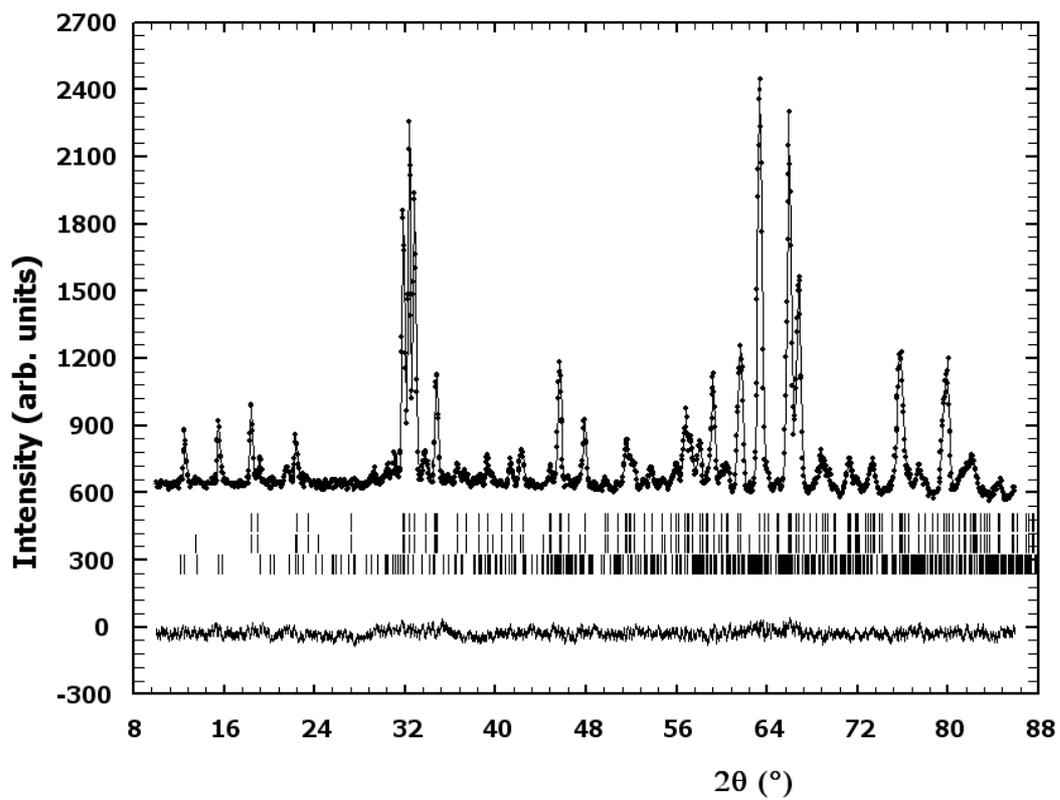

**Figure 8**

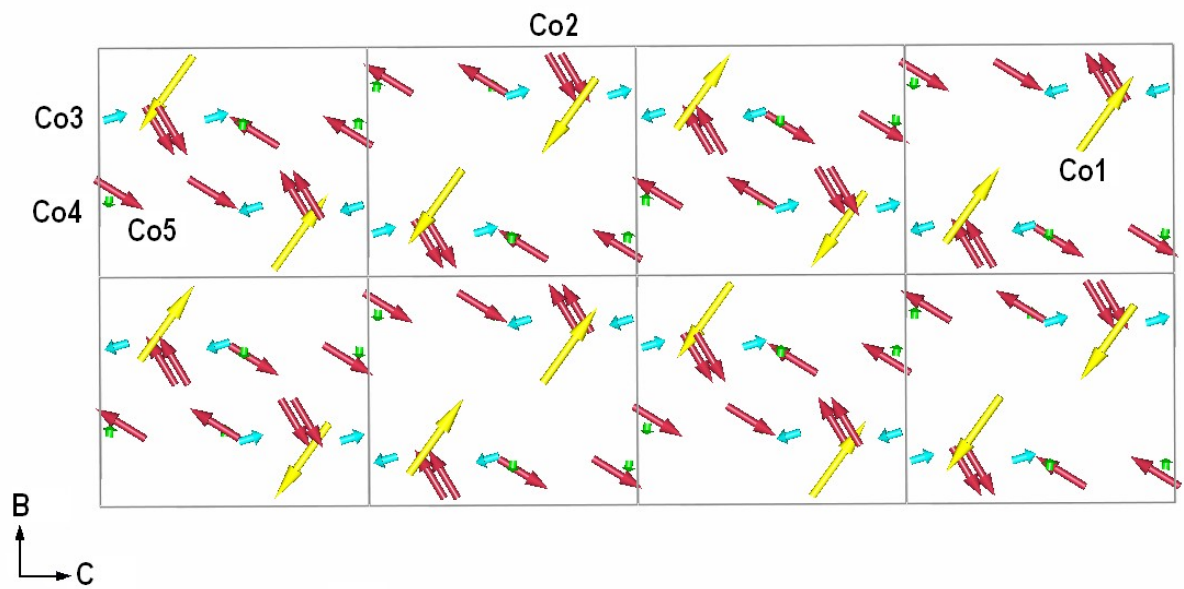

**Figure 9**

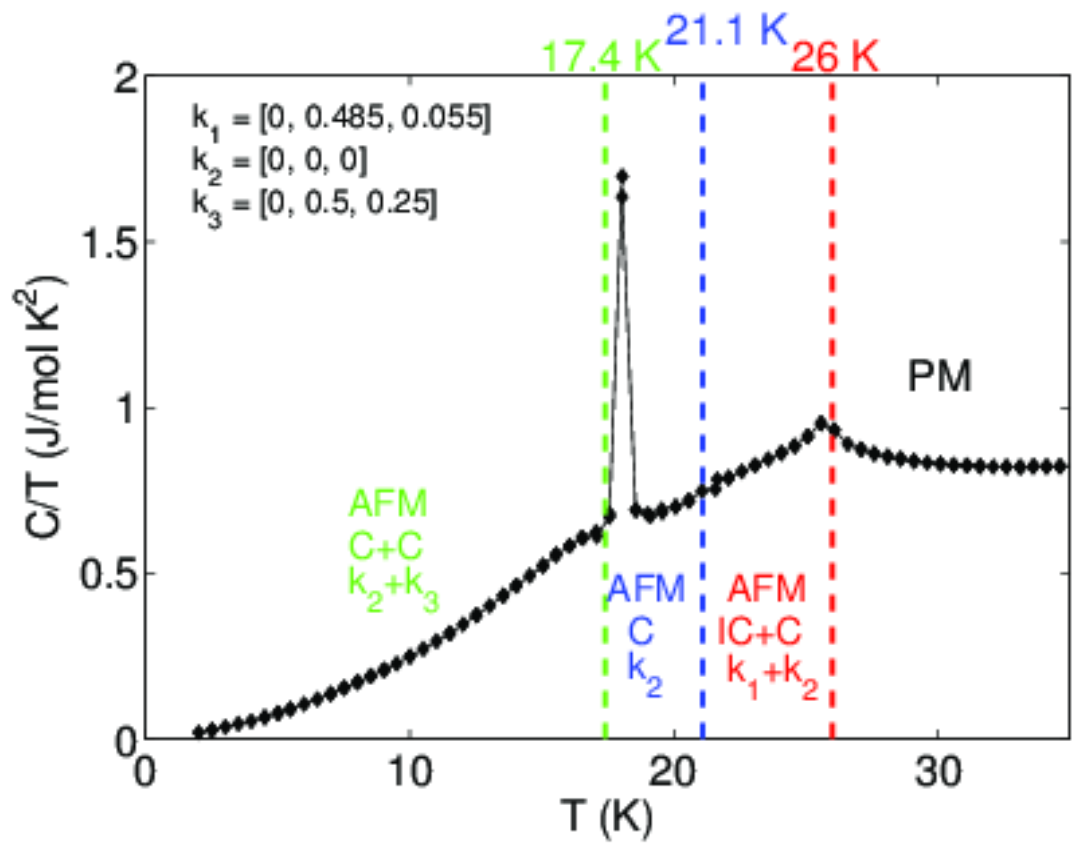

**Figure 10**

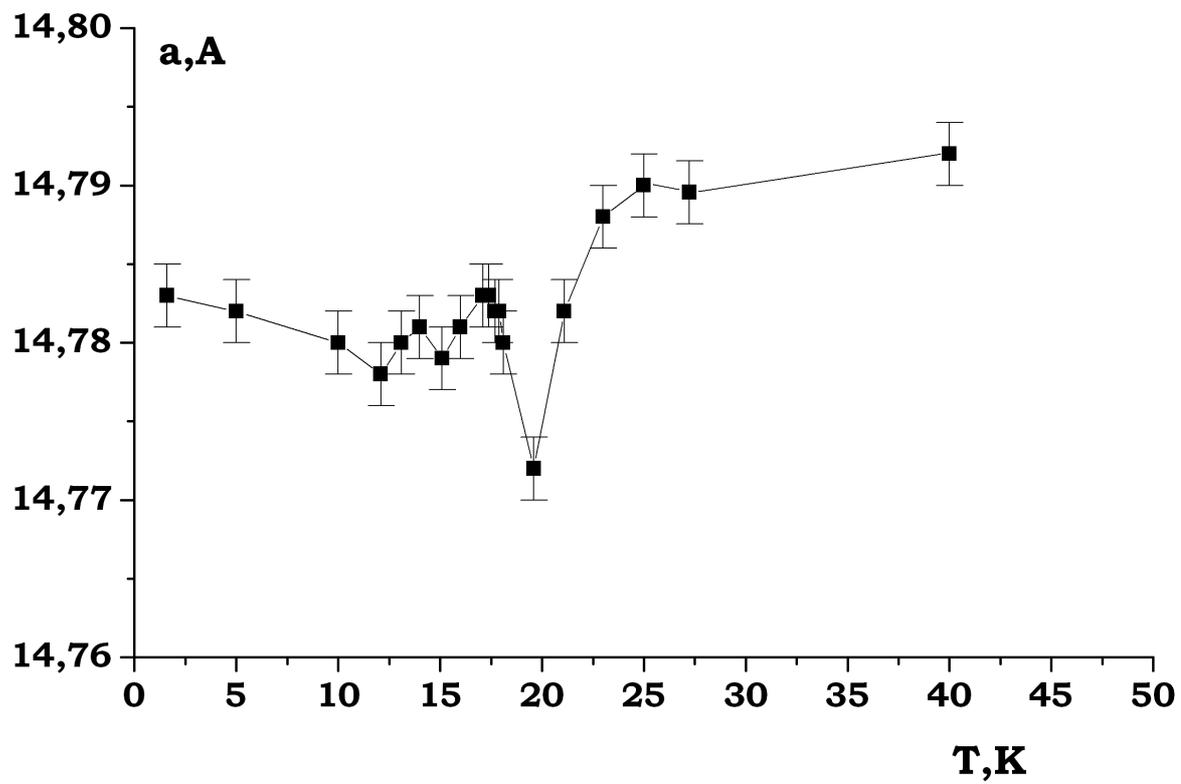

**Figure 11a**

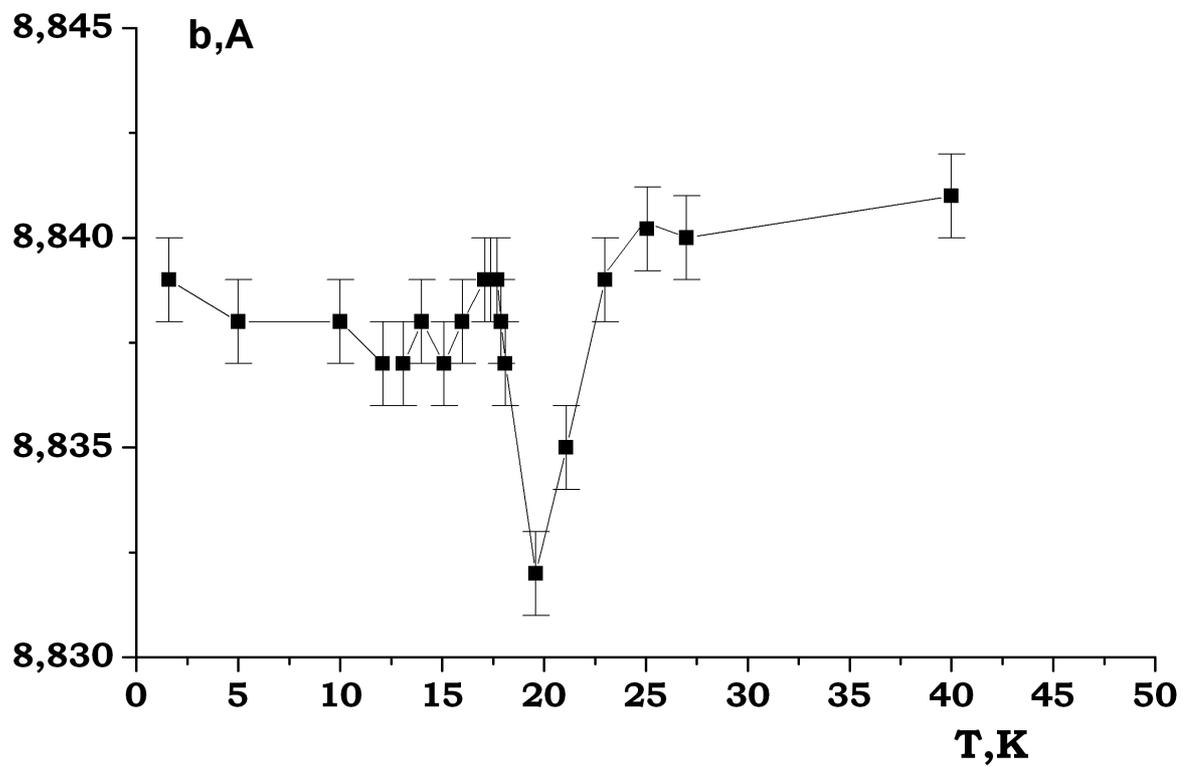

**Figure 11b**

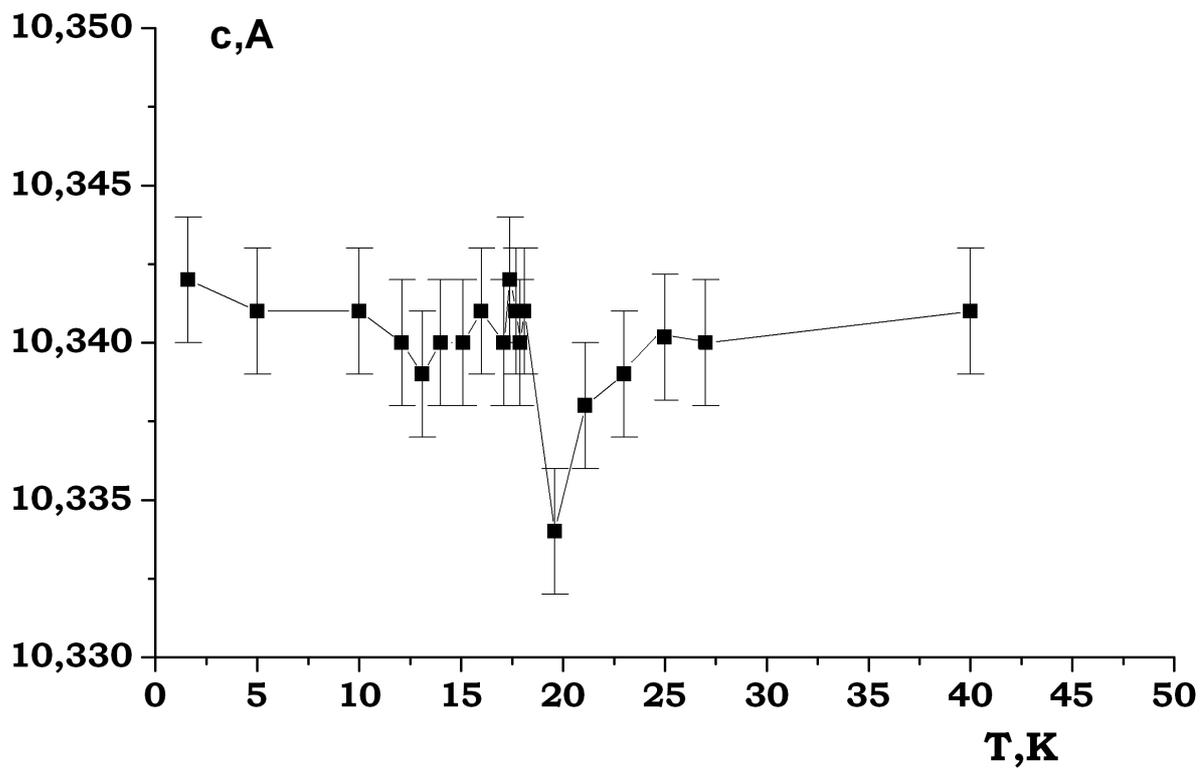

Figure 11c

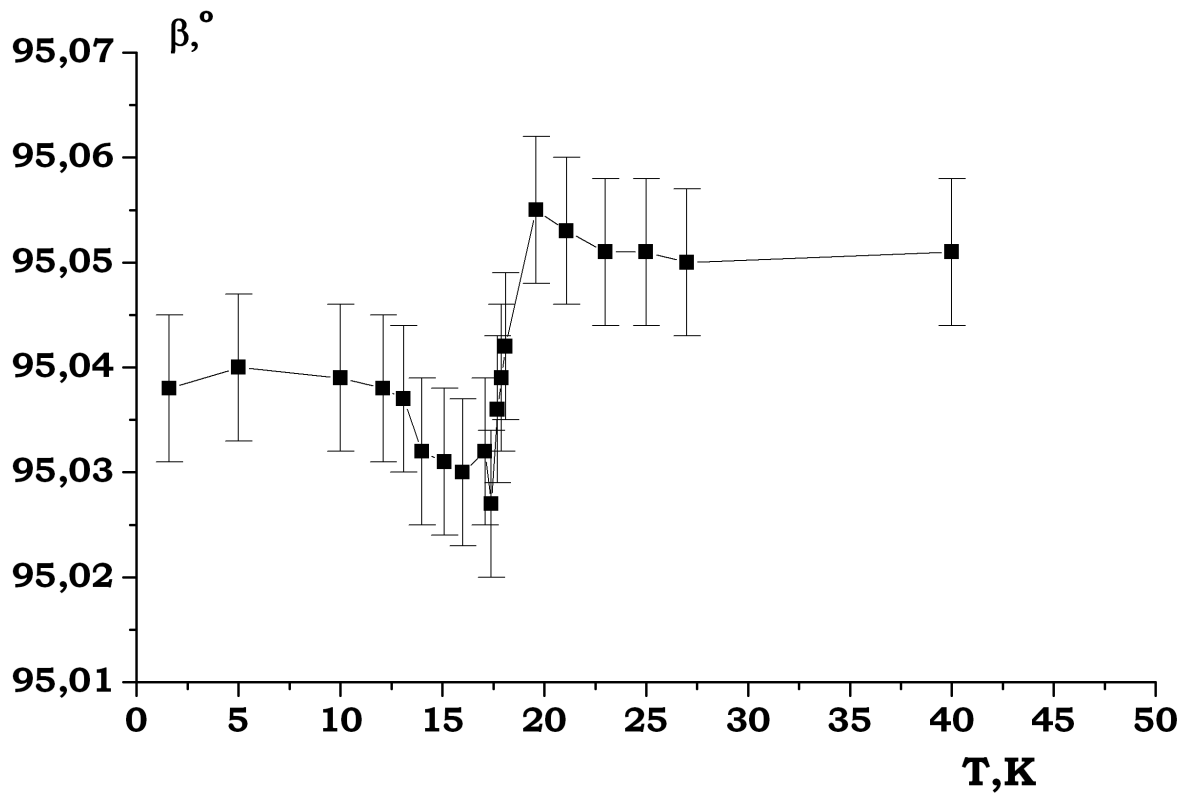

Figure 11d

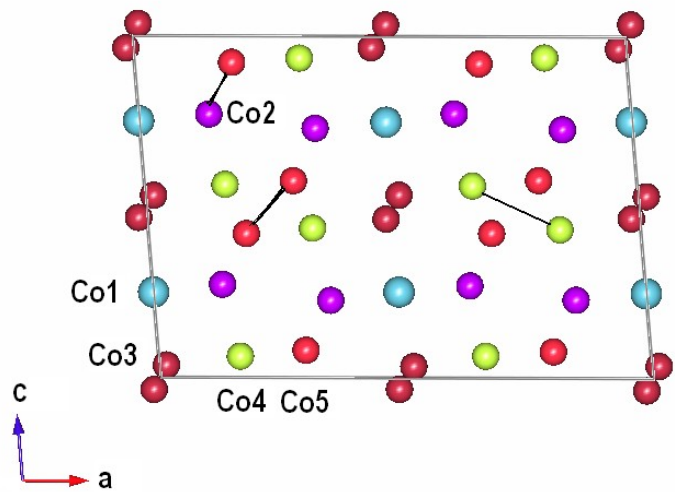

**Figure 12**

**Figure captions**

**Figure 1** Temperature dependence of the zero-field cooled (ZFC) and field-cooled (FC) magnetization in 20 Oe magnetic field (circles, left axis), and heat capacity C, plotted as C/T (diamonds, right axis).

**Figure 2** The observed, calculated, and difference plots for the fit to the XRPD pattern of $Co_3TeO_6$ after Rietveld refinement of the crystal structure at 295K.

**Figure 3** The observed, calculated, and difference plots for the fit to the NPD pattern of $Co_3TeO_6$ after Rietveld refinement of the nuclear structure at 295K (a) and 5K (b).

**Figure 4a** Polyhedral representation of the crystal structure of $Co_3TeO_6$.

**Figure 4b** Projection of a sheet containing Te(1), Co(1) and Co(3) octahedra along [001] in $Co_3TeO_6$. Te(1)-O bonds are emphasized.

**Figure 4c** Projection of a sheet containing Te(2), Co(2), Co(4) octaherda together with Co(5) tetrahedra along [001] in $Co_3TeO_6$. Te(2)-O bonds are emphasized.

**Figure 5(a-b)** Temperature evolution of NPD patterns of Co3TeO6: LLB data. (a) Full NPD patterns at selected temperatures and (b) low angle part of NPD patterns showing subtle changes of the magnetic intensities at separated temperatures

**Figure 6** Temperature evolution of NPD patterns of $Co_3TeO_6$: ILL data

**Figure 7** Schematic representation of the commensurate magnetic structure for $Co_3TeO_6$.

**Figure 8** The observed, calculated, and difference plots for the fit to the NPD patterns of $Co_3TeO_6$ after Rietveld refinement of the nuclear and magnetic structure at 1.6K.

**Figure 9** Schematic representation of the incommensurate magnetic structure for $Co_3TeO_6$ for 2 and 4 cells along the b and c directions, respectively.

**Figure 10** Electronic phase diagram of $Co_3TeO_6$ superposed on the C/T vs. T data. PM-Paramagnetic State; AFM Antiferromagnetic state, IC-Incommensurate Structure; C-Commensurate Structure.

**Figure 11 (a-d)** Thermal evolution of lattice parameters of $Co_3TeO_6$.

**Figure 12** Schematic representation of five different Co cations in nucler structure of $Co_3TeO_6$ at 1.6 K.